\documentclass[aps,prl,showpacs,twocolumn]{revtex4}
\usepackage[brazil,english]{babel}
\usepackage[T1]{fontenc}
\usepackage{amsmath,amsfonts,amssymb,float,graphicx,epsfig,color,times,bbm}

\begin{document}
\title{Vibration enhanced quantum transport}
\author{F. L. Semi\~ao}
\affiliation{Departamento de F\'isica, Universidade Estadual de Ponta Grossa - Campus Uvaranas, 84030-900 Ponta Grossa, Paran\'a, Brazil}
\author{K. Furuya}
\affiliation{Instituto de F\'isica ``Gleb Wataghin'',Universidade
Estadual de Campinas - UNICAMP, Rua Sergio Buarque de Holanda 777, 13083-859, Campinas, Brazil}
\author{G. J. Milburn}
\affiliation{School of Mathematical and Physical Sciences - University of Queensland, Brisbane, Queensland 4072, Australia}
\begin{abstract}
In this paper, we study the role of collective vibrational motion in the phenomenon of electronic energy transfer (EET) along a chain of coupled electronic dipoles with varying excitation frequencies. Previous experimental work on EET in conjugated polymer samples has suggested that the common structural framework of the macromolecule introduces correlations in the energy gap fluctuations which cause coherent EET. Inspired by these results, we present a simple model in which a driven nanomechanical resonator mode modulates the excitation energy of coupled quantum dots and find that this can indeed lead to an enhancement in the transport of excitations across the quantum network. Disorder of the on-site energies is a key requirement for this to occur. We also show that in this solid state system phase information is partially retained in the transfer process, as experimentally demonstrated in conjugated polymer samples. Consequently, this mechanism of vibration enhanced quantum transport might find applications in quantum information transfer of qubit states or entanglement.
\end{abstract}
\pacs{05.60.Gg,78.67.Hc,07.10.Cm,71.35.-y}
\maketitle
\section{Introduction}
The rapid and efficient  transfer of optical excitation energy through a network of coupled dipoles has recently become the subject of intense study in a number of natural and artificial systems.  A typical naturally occurring example is the phenomenon of electronic energy transfer (EET). Many natural photosynthetic proteins and artificial polymers possess light-absorbing molecular dipoles (chromophores) which interact with each other causing the transfer of energy from an initially excited donor
to another acceptor chromophore \cite{scholes}. Artificial systems of coupled semiconductor quantum dots can exhibit similar energy transfer\cite{Curutchet,QD-array}.  The proper
understanding of EET is important for the study of complex quantum systems and may lead to
the design of more efficient devices for energy storage and conversion.

A number of papers in which the authors have tried to understand the basic mechanisms involved in efficient EET have lately appeared in the literature \cite{castro,mohseni,rebentrost,rebentrost2,martin1,m1,m2,martin2,eckel}. The role of quantum coherence in the efficiency of excitation transfer in coupled systems is studied in \cite{castro}. The role of decoherence processes on energy transfer within photosynthetic systems were investigated in \cite{mohseni,martin1} where noise-assisted quantum transport is introduced.  The idea that transport of excitations across chromophoric networks can be enhanced by local dephasing noise was further explored in \cite{rebentrost,rebentrost2,martin2} where it is applied to the description of excitation transport across the Fenna-Matthew-Olson (FMO) complex \cite{fmo}. This system acts like a 'quantum wire' connecting excitations in an antenna complex, where light is absorbed and an exciton created,  to another complex that contains the reaction center where the exciton energy is used to initiate photo-synthetic chemical processes. These studies suggest how nature might use noise-assisted processes to increase the efficiency in the performance of certain tasks. These concepts have recently been extended to the non-Markovian regime\cite{rebentrost3}. An important experiment on EET has recently reported an investigation of the intermediate coupling regime \cite{cs}. This regime interpolates between the classical incoherent hopping (F\"{o}rster) mechanism and a coherent (wavelike) quantum mechanism. The intermediate coupling regime is particularly interesting because both population {\em and} phase information is supposed to be partially transferred through space. Needless to say, such a type of coherent EET would be very interesting for applications in the area of quantum information.

In \cite{cs}, coherent intrachain EET in conjugated polymer samples are studied using specially designed experimental techniques (two-time anisotropy decay  (TTAD) and 2D photon echo experiments) that enable the measurement of coherent EET directly and provide information about coherences. Their results show evidence of coherent EET in the intermediate regime (phase information is partially preserved). Provided that fluctuations of the electronic transition frequencies of the donor and acceptor chromophores are correlated, coherent EET can be observed in two-time anisotropy
decay measurements \cite{cs}. On the contrary, if these fluctuations are uncorrelated, coherent
EET cannot compete with decoherence and the only mechanism left for enhancement of EET seems to be dephasing as previously studied in \cite{mohseni,rebentrost,martin1,martin2}. The experimental results show that for chromophores connected
by the conjugated polymer backbone there are indeed correlated fluctuations in the electronic transition frequencies detected by TTAD. As the authors in \cite{cs} point out, these correlations imply that coherent coupling of the electronic degrees of freedom of the chromophores with the vibrational motion, characteristic of the polymer backbone, must play an important role. In fact, they conjecture that the common
structural framework of the macromolecule is the agent that introduces the correlations that, in turn, preserves the coherence and permits coherent EET.

Of particular relevance to this paper is the recent study of Perdomo et al.,\cite{Perdomo}. They consider an engineered semiconductor quantum dot array and show that carefully controlling the spectral density of the phonon environment can both enhance and suppress energy transfer. In this paper we show how even a single collective vibrational mode of a nanomechanical resonator can be used to achieve the same effect. Our model has the advantage that it can, in principle, be tested in the laboratory under well-controlled conditions using systems such as nanomechanical resonators coupled to quantum dots \cite{Wilson-Rae} or nitrogen-vacancy centers in diamond \cite{xu} or even trapped ions within optical cavities in the weak-coupling regime \cite{cions}. This is essentially the quantum simulator approach \cite{qs}, but it may also provide a novel technique for controlling quantum transport in future engineered multi-component quantum systems.  We do not wish to suggest that this is in any way a good model for complex naturally occurring EET systems, such as the FMO complex or conjugated polymers, although clearly our motivation derives from such chromophoric molecules. 

Our paper is then concerned with
configurational changes in EET which is an old topic in chemical physics \cite{jortner}. Many papers have explicitly included the
vibrational degrees of freedom of biomolecules and studied their effect on varied phenomena. However, most of them are
concerned with the coupling to a broad range of different vibration modes through the embedding environment. On the
other hand, the solid state analogue discussed here makes use of the internal vibronic structure in the spirit of the study
presented in \cite{roden}.  

%
\section{Model}
\begin{figure}[h]
\begin{center}
\includegraphics[scale=0.4]{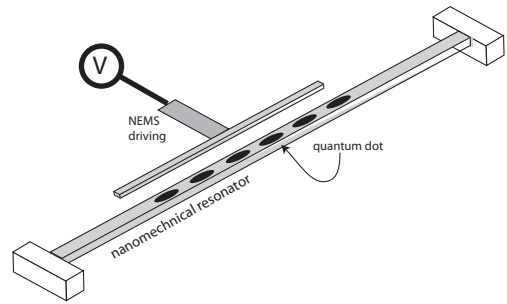}
\end{center}
\caption{One possible realization of the model in which an array of excitonic quantum dots are fabricated inside a a GaAs nanomechanical resonator.  The resonator is driven on resonance by a classical driving field, and is damped. At very low temperatures, the beam is driven to a steady state which is close to a coherent state. The strain induced by the deformation of the nanomechanical resonator shifts the energies of each dot \cite{Wilson-Rae}.  The corresponding physical parameters are given in the appendix.  }
\label{setup}
\end{figure}

We start the discussion of our model by considering a typical setup used to study the transport of energy between two-level dipole systems. It consists of a network of $N$ sites that can exchange excitations by dipole-dipole coupling \cite{mohseni,rebentrost,martin1,martin2}
\begin{eqnarray}
H_N=\sum_{j=1}^N\frac{\omega_{j}}{2}\sigma_z^{j}+\sum_{j\neq k}^N\lambda_{jk}(\sigma_+^{k}\sigma_-^{j}+\sigma_-^{k}\sigma_+^{j}).\label{hn}
\end{eqnarray}
We now suppose that the dipoles are coupled to a {\em single} vibrational mode. This might be realized for example by coupling quantum dots to a single flexural mode of a nanomechanical resonator \cite{Schwab} in such a way that the vibration can modulate the on-site energies of the quantum dots. A specific model of such a system is shown in Fig.(\ref{setup}) where a linear array of coupled excitonic quantum dots are situated in a doubly clamped GaAS nanomechanical resonator.  As shown  by Wilson-Rae \emph{et al}. \cite{Wilson-Rae} for the case of a single quantum dot, the vibrational motion of the beam can indeed change the excitonic energy transition frequency. In the case of a coupled array of quantum dots, all coupled to the same vibrational mode, the modulation in the energy levels will be highly correlated,  but at the same time fluctuations in the resonator will induce dephasing on the coherent hopping spoiling EET as discussed in \cite{roden}.  We will allow the nanomechanical resonator to be classically driven and thus describe it using an underdamped simple harmonic oscillator at non zero temperature.  Again we stress that this is not intended to be a model for natural
molecular EET systems, which are unlikely to be coherently driven; rather they are incoherently excited by thermal motion. However, our model does necessarily include correlated fluctuations in the electronic system due to vibrational motion which is a central idea for EET emerging from studies of molecular systems. 

The dynamics of the model, including  dissipation, is described by a master equation for the total density operator $\zeta$ (quantum dots plus nanomechanical resonator) \cite{GZ} 
\begin{eqnarray}
\frac{d \zeta}{d t}=-i[H_{NV},\zeta]+\gamma(\bar{n}+1)\mathcal{D}[a]\zeta+\gamma \bar{n}\mathcal{D}[a^\dagger]\zeta,\label{m1}
\end{eqnarray}
with
\begin{eqnarray}
H_{NV}=H_N+\nu \hat{a}^\dag \hat{a}+\varepsilon (a^\dag e^{-i\nu t}+ a e^{i\nu t})+\hat{q}\sum_{j=1}^Ng_j\sigma_z^{j},\label{hall}
\end{eqnarray}
where $H_N$ given by Eq.(\ref{hn}), $\varepsilon$ is the amplitude of driving force on the vibrational mode, and $g_j$ is the coupling constant between the site $j$ and the motional displacement operator $\hat{q}=q_0(\hat{a}^\dag+\hat{a})$.  This coupling generates correlated frequency fluctuations on the sites as the nanomechanical oscillator vibrates. A very interesting proposal for generating a non driven version of $H_{NV}$ in the context of trapped ions has recently appeared \cite{Giorgi}. In Eq.(\ref{m1}) there is also a term that describes the vibrational damping at rate $\gamma$, where $\bar{n}$ is the mean thermal occupation of the environment seen by the oscillator,  and $\mathcal{D}[a]$ is a super-operator defined as $\mathcal{D}[A]B=ABA^\dag-\{A^\dag A,B\}/2$. We are using the quantum optics master equation, Eq.(\ref{m1}), which assumes that the mechanical motion is underdamped. This has been used successfully to describe recent experiments in nanomechanics \cite{wilson-rae-cooling} at low temperature. In the much more complicated case of molecular systems there are many vibrational modes and a quite different approach is required. 

Much like dephasing enhanced transfer \cite{mohseni,rebentrost,martin1,martin2}, vibration enhanced transfer does not occur for all possible Hamiltonians of the type given by (\ref{hall}), i.e. for arbitrary $\omega_j$, $\lambda_{jk}$, and $g_j$. In order to present an example of vibration enhanced transfer, we consider a simple, yet illustrative, case of a linear chain with only nearest neighbor interactions with equal coupling constants. In this case, the system Hamiltonian reads
\begin{eqnarray}
H_{NV}&=&\sum_{j=1}^N\frac{\omega_{j}}{2}\sigma_z^{j}+\nu \hat{a}^\dag \hat{a}+\lambda\sum_j^N(\sigma_+^{j}\sigma_-^{j+1}+\sigma_-^{j}\sigma_+^{j+1})\nonumber\\ &&+\varepsilon (a^\dag e^{-i\nu t}+ a e^{i\nu t})+q_0(\hat{a}^\dag+\hat{a})\Sigma_z.\label{hl}
\end{eqnarray}
with $\sigma_+^{N+1}=0$ and the collective operator $\Sigma_z=\sum_{j=1}^{N} g_j\sigma_z^j$. The vibrational motion of the nanomechanical resonator is damped and the driving will settle it into a stable steady state oscillation at the driving frequency. For a very low temperature of the cooled resonator, this steady state will be close to a coherent state, but subject to zero point fluctuations. These fluctuations act as a dephasing environment for the quantum dots.  At finite temperature the steady state is a displaced thermal state and the dephasing environment is thermal.   

We transform the system Hamiltonian to an interaction picture at the oscillator frequency and displace the bosonic operators $a,a^\dagger$ to subtract the coherent steady state values they would have in the absence of coupling to the quantum dots, enabling us to treat the steady state of the vibrational motion as purely thermal. To this end, we first apply the displacement operator $D(\beta)=\exp(\beta a^\dag-\beta^* a)$, with $\beta=-i \beta_0 = - 2i\varepsilon/\gamma$, such 
that the master equation (\ref{m1}) will assume the form [$\tilde{\rho}=D(-\beta)U_1^\dag(t)\zeta U_1(t) D(\beta)$]
\begin{equation}
\frac{\partial \tilde{\rho}}{\partial t}= \mathcal{L}_2\tilde{\rho}+{\cal L}_c\tilde{\rho}+{\cal L}_1\tilde{\rho},
\label{master2}
\end{equation}
where
\begin{eqnarray}
\mathcal{L}_2\tilde{\rho}&=&-i[\sum_{j=1}^N\frac{\omega_{j}}{2}\sigma_z^{j}+\lambda\sum_j^N(\sigma_+^{j}\sigma_-^{j+1}+\sigma_-^{j}\sigma_+^{j+1}),\tilde{\rho}]\nonumber\\ &&
- 2q_0\beta_0 \sin \nu t\,[\Sigma_z,\tilde{\rho}]\\
\mathcal{L}_c\tilde{\rho}&=&-iq_0[(a^\dagger e^{i\nu t}+ae^{-i\nu t})\Sigma_z,\tilde{\rho}]\\
{\cal L}_1\tilde{\rho} & = & \gamma(\bar{n}+1) \mathcal{D}[a]\tilde{\rho}+\gamma\bar{n} \mathcal{D}[a^\dagger]\tilde{\rho}
\end{eqnarray}
In physical terms $|\beta| q_0$ is of the same order as the maximum coherent displacement of the driven nanomechanical resonator. Equivalently $|\beta|^2$ is the average phonon number excitation of the nanomechanical resonator in the steady state.  

In the case of EET in molecular systems, the electronic degrees of freedom are coupled to many vibrational modes. In that case, the description usually proceeds by treating the vibrational motion as some kind of heat bath leading to dephasing of the electronic excitations. In the simplest model this is treated as if each site was coupled to an independent heat bath \cite{haken,yang,mohseni,martin1}, but in recent years it has become clear that this is not a good model for real light harvesting systems and temporal and spatial correlations in the dephasing environments arise due to the common vibrational motion of the protein cage in which the chromophores are situated \cite{cs,fmo,lee,sarovar,nazir,Fassioli}. In the simpler physical context of this paper there is only one vibrational degree of freedom, which can hardly be treated as a heat bath. However we can make contact with other models by assuming that the vibrational motion is rapidly damped on a time scale of the dipole-dipole coupling and the dipole-vibrational coupling. 

If the vibrational degree of freedom relaxes rapidly to a steady state on the time scale for energy transfer through dipole-dipole coupling $(\lambda^{-1})$, we can adiabatically eliminate this motion \cite{WM,elimination}. Notice that if $g_j=0$ for all $j$, the state of the oscillator would be a coherently displaced thermal state in the limit $t>>\gamma^{-1}$ (equilibrium state). In the adiabatic limit, the excitonic systems will thus respond as if it were coupled to a heath bath; however we need to take care not to average away the coherent driving of the mechanical motion in our model.  We use a version of the Zwanzig projection operator method due to Gardiner and Eschmann \cite{Gardiner}. 
The Zwanzig projection operator method then gives the following equation for the quantum dots alone, 
\begin{equation}
  \dot{\sigma} = \mathcal{L}_2\sigma + \int_0^\infty
  dt' {\rm Tr}_1 [ \mathcal{L}_c(t)\exp (\mathcal{L}_1 t')
  \mathcal{L}_c(t-t') (\rho_{\rm ss} \otimes \sigma)].
\end{equation}
where $\rho_{ss}$ is the steady state of the vibrational mode in the absence of the interaction with the quantum dots. 
This last expression allows us to find the master equation
for the quantum dot system in terms of timecorrelation
functions of the vibrational system evolving under $\mathcal{L}_1$.
We find here that
\begin{eqnarray}
\frac{\partial \sigma}{\partial t}= -i [H_{e},\sigma]+ 4\Gamma \mathcal{D}[\Sigma_z]\sigma,
\label{master4}
\end{eqnarray}
where $\Gamma=q_0^2(2\bar{n}+1)/\gamma$ and $H_{e}$ is the effective Hamiltonian for the network of quantum dots which is given by
\begin{equation}
H_{e}= \sum_{j=1}^{N} \chi_j(t) \sigma_z^j+ 
\lambda \sum_{j=1}^{N}(\sigma_+^{j}\sigma_-^{j+1} + \sigma_+^{j+1}\sigma_-^{j}),
\label{Hef}
\end{equation}
with the time-dependent coefficients $\chi_j(t)= \omega_j/2-2g_j \beta_0 q_0\sin \nu t$.
We can see that the coupling with a driven and heavily damped oscillator not only introduces frequency modulations but also {\em collective} dephasing \cite{note-1}. The role played by the external drive on the nanomechanical resonator is now evident. It allows the control of $\chi_j(t)$ through the parameter $\beta_0$ which is a function of the externally controlled amplitude driving $\varepsilon$ and the damping $\gamma$. Our goal in next section is to show that by varying $\beta_0$, it is possible to find more favorable environments for EET when compared to the case of no drive, $\beta_0=0$. All simulations in this paper will be performed using Eq.(\ref{master4}). 
\section{Results}
In order to find how efficiently an excitation injected at one end of the chain (donor) reaches the other end (acceptor) we include an irreversible mechanism of energy loss (sink) in the acceptor at the site $j = N$. By considering only conditional evolutions where no absorptions occurred up to time $t$, we can evaluate the rate of energy transfer as the emission rate into the sink $R(t)=2\kappa\langle\Psi_c(t)|\sigma_+^N\sigma_-^N|\Psi_c(t)\rangle$ \cite{qj} . This can be easily done by adding a non Hermitian term $-i\kappa|e\rangle_N\langle e|$ to Hamiltonian (\ref{Hef}) and evaluating the time evolved state $|\Psi_c(t)\rangle$ where the index $c$ reminds us that this is evolution under the condition of non-absorption. From this rate, the efficiency for a fixed time interval $t$ is then calculated as
\begin{eqnarray}
\mathrm{Efficiency}=\int_0^t R(t^{'})dt^{'}.
\end{eqnarray}
Now we define the basis $\mathcal{B}=\{|ggg...g\rangle=|0\rangle, |egg...g\rangle=|1\rangle,|geg...g\rangle=|2\rangle,|gge...g\rangle=|3\rangle,...|ggg...e\rangle=|N\rangle \}$ which is clearly invariant under the action of $H_{e}$. The density operator in this basis is then written as $\sigma=\sum_{i,j=0}^{N}\sigma_{ij}|i\rangle \langle j |$, and the set of coupled differential equations for its matrix elements $\sigma_{ij}$ obtained from (\ref{master4}) is given by
\begin{eqnarray}
\dot{\sigma}_{ij}&=&\{-2i(\chi_i(t)-\chi_j(t))+4 \Gamma G_{ij}-\kappa[\delta_{jN}+\delta_{iN}]\}\sigma_{ij}\nonumber\\ 
 &&-i\lambda [F_{i1}\sigma_{i-1j}+F_{iN}\sigma_{i+1j}- F_{j1}\sigma_{ij-1}\nonumber\\ &&-F_{jN}\sigma_{ij+1}],
\label{eqmotion}
\end{eqnarray}
in the case  $i,j\neq 0$ where $F_{ij}=1-\delta_{ij}$, and 
\begin{eqnarray}
\dot{\sigma}_{0j}&=&[2i \chi_j(t)+4 \Gamma G_{0j}-\kappa\delta_{jN}]\sigma_{0j}\nonumber\\ 
 && + i\lambda F_{j1}\sigma_{0j-1}+i\lambda F_{jN}\sigma_{0j+1},
\label{eqmotion0}
\end{eqnarray}
for the elements associated with the state $|0\rangle$, where $G_{ij}=-2(g_i-g_j)^2$ with $g_0=0$.

\begin{figure}
 \centering\includegraphics[width=\columnwidth]{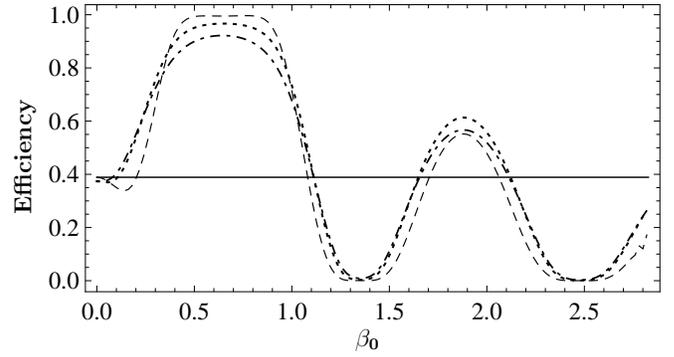}
 \caption{Efficiency as a function of $\beta_0$ for an integration time $t=3000\lambda$. The network frequencies are $\omega_3=1.0$ and $\omega_1=\omega_2=\omega_4=\omega_5=\omega_6=0$, the couplings between quantum dots and the nanomechanical resonator $g_3=1.5$ and $g_1=g_2=g_4=g_5=g_6=0.5$, decay constant to the sink $\kappa=0.2$, mean number of thermal phonons $\bar{n}=5$, and inter-quantum dots coupling $\lambda=0.1$. The different curves correspond to $\gamma$ equals to $1.1\times 10^5$ (dashed), $1.1\times 10^3$ (dotted), and $5.5\times 10^2$ (dot-dashed).}
 \label{fig1}
\end{figure}

\begin{figure}
 \centering\includegraphics[width=\columnwidth]{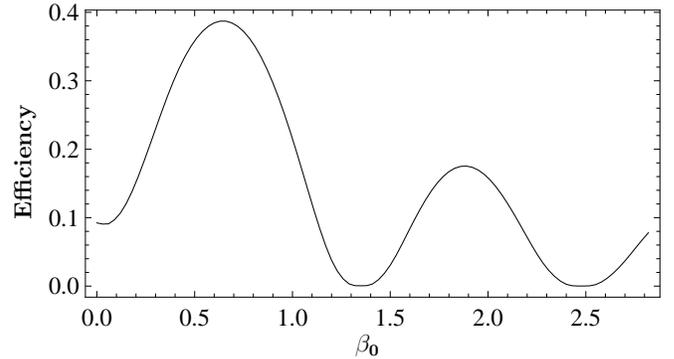}
 \caption{Averaged efficiency over $1000$ realizations as a function of $\beta_0$ for an integration time $t=3000\lambda$. The network random frequencies are chosen from six independent Gaussian distributions having mean values $\omega_3=1.0$ and $\omega_1=\omega_2=\omega_4=\omega_5=\omega_6=0$, and standard deviation $0.1$, respectively. Decay constant $\gamma$ is chosen as $\gamma=1.1\times 10^5$, and the couplings between quantum dots and nanomechanical resonator as well as other parameters are the same as in Fig.(\ref{fig1}).}
 \label{fig2}
\end{figure}

\begin{figure}
 \centering\includegraphics[width=\columnwidth]{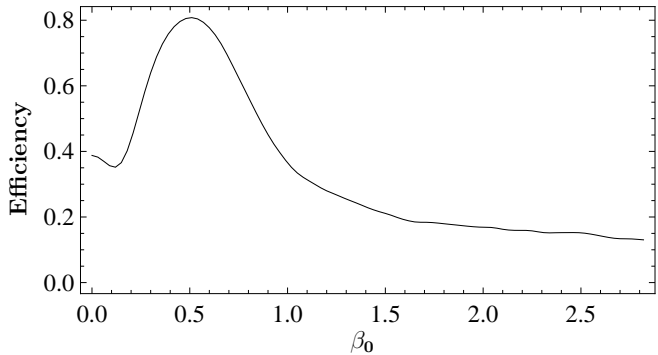}
 \caption{Averaged efficiency over $1000$ realizations as a function of $\beta_0$ for an integration time $t=3000\lambda$. The couplings between the quantum dots and the nanomechanical resonator are chosen from six independent Gaussian distributions having mean values $g_3=1.5$ and $g_1=g_2=g_4=g_5=g_6=0.5$, and standard deviation $0.3$, respectively. Decay constant $\gamma$ is chosen as $\gamma=1.1\times 10^5$, and the quantum dots' frequencies and other parameters are the same as in Fig.(\ref{fig1}).}
 \label{fig3}
\end{figure}

Using the above equations we numerically obtain the efficiency for $N=6$ quantum dots using arbitrary units $m=\hbar=\nu=1$ (see the appendix for the physical parameters pertinent to a physical example). Please notice that this leads to $q_0=1/\sqrt{2}$. In Fig.(\ref{fig1}), we can clearly see that there are amplitudes $\beta_0$ leading to an improvement on the efficiency when compared to the case with no oscillator [constant curve (solid)]. 
In order to analyze the robustness of this mechanism of EET, we have also performed a couple of simulations considering either $\omega_j$ or $g_j$ as random variables. First, we consider the same values of $g_j$ used in Fig.\ref{fig1}, but now the quantum dots' frequencies are chosen from independent Gaussian distributions. In Fig.(\ref{fig2}), we show the averaged efficiency over 1000 realizations. When compared to the Fig.(\ref{fig1}), one can see that the effect of this randomness on the  frequencies is to decrease the averaged efficiency. It is interesting though that there are still values of $\beta_0\neq 0$ leading to an improvement of efficiency. Finally, we now maintain $\omega_j$ fixed as in Fig.(\ref{fig1}), but we choose the couplings  as random variables again according to independent Gaussian Distributions. The result is shown in Fig.(\ref{fig3}). The randomness not only decreases the maximum efficiency but also tends to restrict the values of $\beta_0$ leading to an enhancement of EET by narrowing the peak. Nonetheless,  there are still many values of $\beta_0$ for which EET is favored. These results  indicate that vibration enhanced EET is effective even in the presence of slight randomness. 

\begin{figure}[H]
 \centering\includegraphics[width=\columnwidth]{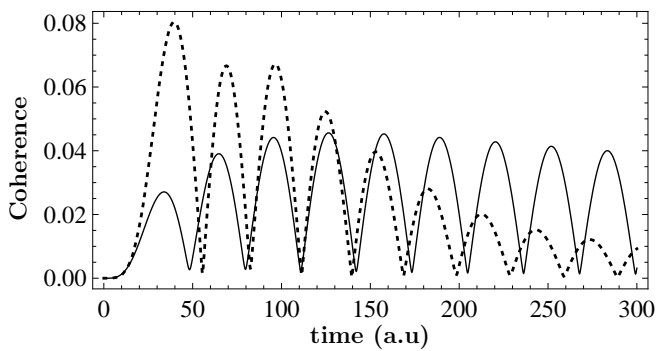}
 \caption{Time evolution of coherence for the initial state $|\phi(0)\rangle=(|e\rangle_1+|g\rangle_1)|ggggg\rangle_{23456}/\sqrt{2}$. The network frequencies are $\omega_3=1.0$ and $\omega_1=\omega_2=\omega_4=\omega_5=\omega_6=0$, couplings between quantum dots and vibration mode $g_1=g_2=g_3=g_4=g_5=g_6=0.0$ (solid), i.e. absence of vibration, $g_3=1.5$ and $g_1=g_2=g_4=g_5=g_6=0.5$ (dotted). The decay constant to the sink is $\kappa=0.2$, decay constant of the oscillator $\gamma=1100$, $\beta_0=0.65$, mean number of thermal phonons in the environment $\bar{n}=5$, and inter-quantum dots coupling $\lambda=0.1$.}
 \label{fig4}
\end{figure}

If we trace out the first five quantum dots, the resulting density matrix of the sixth quantum dot will have the off-diagonal element given by $\sigma_{06}(t)$. In order to investigate whether or not the coupling to the vibrational degree of freedom can partially preserve the coherence of EET, we have calculated $|\sigma_{06}(t)|$ for the initial state $|\phi(0)\rangle=(|e\rangle_1+|g\rangle_1)|ggggg\rangle_{23456}/\sqrt{2}$. The result is shown in Fig.(\ref{fig4}). It can be seen that, for short times, the presence of the oscillator (dotted) does indeed improve the coherence when compared to the case with no oscillator (solid). In this plot, we have kept the sink coupled to the last quantum dot $\kappa\neq 0$. Otherwise, the coherence would assume much higher values.
This is an example of enhancement of quantum transport in a network of two-level systems by coherent coupling with a driven resonator. The role of incoherent vibronic coupling (phonon-induced decoherence) in EET has already been investigated \cite{rebentrost2,roden}. Our simplified model, valid in the intermediate coupling, has shown the possibility of coherent EET with partial preservation of phase information, something unlikely to take place in noise assisted EET.  

\section{Discussion and conclusions}
As one can see from Fig.\ref{fig1}, energy transport occurs at particular values of the mechanical resonator driving strength. To understand this we neglect the decoherence due to the ground state fluctuations of the vibrational mode as well as absorption at the special site and consider only the network Hamiltonian which we write as
\begin{equation}
H=\sum_{j=1}^N\frac{\omega_j(t)}{2}\sigma_z^j+\lambda\sum_{j=1}^N(\sigma_+^j\sigma_-^{j+1}+\sigma_+^{j+1}\sigma_-^j),
\end{equation}
where the modulated site energies are defined as
\begin{equation}
\omega_j(t) = \omega_j-4g_j q_0 \beta_0 \sin\nu t.
\label{mod-energy}
\end{equation}

We now transform to an interaction picture via
\begin{equation}
U_0(t)=\exp\left [-\frac{i}{2}\sum_{j=1}^N\int_0^t dt'\omega_j(t')\sigma_z^j\right],
\end{equation}
to obtain the interaction Hamiltonian
\begin{equation}
H_I(t)=\lambda\sum_{j=1}^N\ [\sigma_+^j\sigma_-^{j+1} e^{i\theta(t)}+h.c.],
\end{equation}
where $\theta(t)=\Delta\omega_j t-4i\Delta g_j(\beta_0 q_0/\nu)\left (\cos\nu t-1\right )$, $\Delta\omega_j=\omega_j-\omega_{j+1}$ and $\Delta g_j=g_j-g_{j+1}$.  Expanding the oscillating exponential as a Fourier series using the generating function for Bessel functions $J_n(x)$, we obtain
\begin{equation}
H_I(t) = \lambda\sum_{n=-\infty}^{n=\infty} \sum_{j=1}^N[\mathcal{J}_n e^{i(\Delta\omega_j-n\nu)t}\sigma_+^j\sigma_-^{j+1}+h.c.],
\label{int-ham-res}
\end{equation}
where we defined the coefficients
\begin{eqnarray}
\mathcal{J}_n= (i)^{-n}J_n(4\Delta g_j\beta_0 q_0/\nu)e^{4i\Delta g_j\beta_0 q_0/\nu}.
\end{eqnarray}
It is now apparent that there will be {\em resonances} when 
\begin{equation}
\Delta\omega_j=n\nu.
\end{equation}
We thus expect that when the differences between site frequencies is of the same order as the mechanical vibrational frequency, the resonant terms in Eq.(\ref{int-ham-res}) will dominate the interaction, provided the Bessel function coefficients have sufficient weight at those frequencies. These resonances are essentially due to the sidebands written on the electronic transitions by the phase modulation implicit in  Eq.(\ref{mod-energy}). The Bessel functions are a function of the driving amplitudes, $\beta_0$, and also the coupling inhomogeneities, $\Delta g_j$. Note that if the system is completely homogeneous so that $\Delta\omega_j=0$ and  $\Delta g_j=0$, there are no resonances. Inhomogeneity plays a crucial role.  Given some inhomogeneity, the resonances can be controlled by the external driving strength through the dependence on $\beta_0$.

\begin{figure}[H]
 \centering\includegraphics[width=\columnwidth]{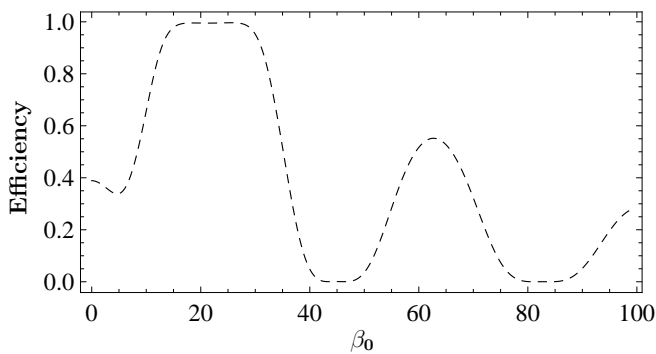}
 \caption{Same as Fig.(\ref{fig1}) but now with $g_3=0.03$ and $g_1=g_2=g_4=g_5=g_6=0$ and equal to $\gamma=1.1\times 10^5$. This value of $g$ is consistent with \cite{Wilson-Rae}, as discussed in the appendix.}
 \label{fig6}
\end{figure}

We can see the effect of the Bessel function weighting on the resonances in Fig \ref{fig1}.  In the dimensionless units of this example the vibrational frequency is $\nu=1$ while the largest difference in site frequencies is $\Delta\omega_2=-\Delta\omega_3=1.0$ so a first order resonance is possible. The first two zeros of the first order Bessel function occur at $x_1=3.83$ and $x_2=7.02$. This would imply a suppression of the resonance for the corresponding values of $\beta_0\sim 1.35$ and $\beta_0\sim 2.5$ and we do indeed see a suppression of EET near these values. Note that it is the product of the driving strength $\beta_0$ and the coupling differences $\Delta g_j$ that matters. If the coupling is weak, one needs higher driving strengths to see the same phenomenon. This is illustrated in Fig.(\ref{fig6}), where the zeros of the first order Bessel function would now imply in suppression of EET at $\beta_0\sim 45$ and $\beta_0\sim 83$, in very good agreement with the numerical plotted curve. Likewise, larger vibrational frequencies require a higher driving strength.
 
To summarize, we have shown that for an inhomogeneous linear chain of coupled quantum dots in a harmonically driven vibrational structure,  particular driving strengths enhance the quantum transport of excitations across the network. Future work may cover the inclusion of additional vibrational degrees of freedom and move beyond nearest neighbor coupling. We have also shown that vibration enhanced quantum transport can partially preserve quantum coherences during electronic energy transfer. This fact may motivate future research on propagation of quantum information in this system.

\emph{Acknowledgments:}
F. L. S. and K.F. wish to thank the Australian Research Council Centre for Quantum Computer Technology at The University of Queensland for hosting their visit to Brisbane.  F. L. S. and K. F. acknowledge participation as members of the Brazilian National Institute of Science and Technology of Quantum Information (INCT/IQ). F.L.S. acknowledges partial support from CNPq under grant 303042/2008-7.  K.F. also acknowledges partial support from CNPq.  G. J. M. would like to acknowledge the support of the Australian Research Council, and to thank Andrew Ringsmuth for useful discussions on the nature of energy transport in FMO and other photosynthetic complexes.
\section{Appendix}
In this appendix we calculate the model parameters for a specific realization based on quantum dots in GaAs, following the proposal of Wilson-Rae et al. \cite{Wilson-Rae}. The coupling between the vibrational mode and a single quantum dot in their model is given by
\begin{equation}
H_{VD}=\frac{\hbar\nu\eta}{2}\sigma_z\hat{q}
\end{equation}
where $\eta$ is a dimensionless parameter given by Eq. (3) in \cite{Wilson-Rae}, which we also take as $\eta=0.06$. Comparing this to Eq.(\ref{hall}) in our model we see that
\begin{equation}
\hbar g=\frac{\hbar\nu\eta}{2q_0}
\end{equation}
where $q_0$ is the rms position in an oscillator ground state and is given by
\begin{equation}
q_0=\left (\frac{\hbar}{2m\nu}\right )^{1/2}
\end{equation}
We use the results from \cite{Wilson-Rae} for a GaAs beam of length, $L=1\ \mu$m, width $b=85\ $nm and depth $d=30\ $nm, the mass is of the order of $m\sim 1.4\times 10^{-17}\ $kg, the fundamental mode frequency is $\nu=1.2\times 10^9\ $s$^{-1}$  and $q_0\sim 5\times 10^{-14}\ $m. This implies that $g\sim 36 \times 10^{6}\ $s$^{-1}$ in standard units. In the units of this paper such value of $g$ translates as $g=\eta/2=0.03$, and this was considered in Fig.(\ref{fig6}).

The damping rate for the nanomechanical resonator is determined by the quality factor, $Q$, of this resonance. If we take the rather conservative value of 
$Q=10^2$, then at a resonance frequency quoted by Wilson-Rae et al.\cite{Wilson-Rae}, we have that $\gamma\sim 10^{7}\ $ s$^{-1}$ leading to $gq_0/\gamma\sim 1$ which is the weak coupling regime. In this regime, the oscillator is heavily damped compared the coherent coupling with the quantum dots. Should this limit not be sufficient to safely eliminate the vibrational motion adiabatically, a lower mechanical resonance frequency or a lower Q resonator may be used.

The site energy for a GaAs quantum dot is of the order of $\hbar\omega_i\sim 1$ meV \cite{NazLov}. From  Perdomo et al.\cite{Perdomo} we take the ratio of the coupling strength to the excitation energy of the quantum dot to be to be of the order of $\frac{\lambda}{\omega_i}\sim 0.1\ $. This corresponds to quantum dots separated by about $10\ $nm. In order to observe the kind of resonance discussed in this paper we need the site energy inhomogeneity, $\Delta (\hbar\omega)$ to be of the order of the vibrational energy quanta $\hbar\nu$. For the case of $\nu= 1.2 \times 10^9$s$^{-1}$, this would imply $\Delta\omega \sim 1 \mu$eV. This is somewhat smaller than typical inhomogeneities that occur due to growth conditions. However external electric fields can be used to tune quantum dots to a considerable degree \cite{Stinaff}, so it may be possible to set up conditions to observe the phenomenon discussed in this paper using coupled InAS/GaAs quantum dots in a nanomechanical resonator. A detailed description and theory of such an experiment will be given in a future paper. 

\emph{Note added in proof}: While being considered for publication, some independent studies
related to this work appeared in the literature \cite{nap}


\begin{thebibliography}{xxxxx}
\bibitem{note-1}Given that this is collective dephasing, the possibility of decoherence free subspaces for the electronic dynamics is something that should be considered in future work.
\end{thebibliography}
\end{document}